\documentclass[12pt,thmsa]{article}
\usepackage{amssymb}

\usepackage{sw20rui}

\input{tcilatex}
\begin{document}

\title{\textbf{Twisting Null Geodesic Congruences, Scri, H-Space and Spin-Angular
Momentum}}
\author{Carlos Kozameh$^{1},$ \and Ezra. T. Newman$^{2}$ \and Gilberto Silva-Ortigoza%
$^{3}$ \\
$^{1}$FaMaF, Univ. of Cordoba, Cordoba, Argentina\\
$^{2}$Dept of Physcis and Astronomy, Univ. of Pittsburgh, Pittsburgh, PA
15260, USA\\
$^{3}$Facultad de Ciencias F\'{\i }sico Matem\'{a}ticas de la Universidad
Aut\'{o}nomade Puebla, \\
Apartado Postal 1152, Puebla, Pue.,Mexico}
\date{June 8,.2005}
\maketitle

\begin{abstract}
The purpose of this work is to return, with a new observation and rather
unconventional point of view, to the study of asymptotically flat solutions
of Einstein equations. The essential observation is that from a given
asymptotically flat space-time with a given Bondi shear, one can find (by
integrating a partial differential equation) a class of asymptotically
shear-free (but, in general, twistiing) null geodesic congruences. The class
is uniquely given \textit{up to the arbitrary choice of a complex analytic
world-line in a four-parameter complex space}. Surprisingly this parameter
space turns out to be the $H$-space that is associated with the real
physical space-time under consideration. The main development in this work
is the demonstration of how this complex world-line can be made both unique
and also given a physical meaning. More specifically by forcing or requiring
a certain term in the asymptotic Weyl tensor to vanish, the world-line is
uniquely determined and becomes (by several arguments) identified as the
`complex center-of-mass'. Roughly, its imaginary part becomes identified
with the intrinsic spin-angular momentum while the real part yields the
orbital angular momentum.

One should think of this work as developing a generalization of the
properties of the algebraically special space-times in the sense that the
term that is forced here to vanish, is automatically vanishing (among many
other terms) for all the algebraically special metrics. This is demonstrated
in one of several given examples. It was, in fact, an understanding of the
algebraically special metrics and their associated \textit{shear-free null
congruence }that led us to this construction of the \textit{asymptotically
shear-free} congruences and the unique complex world-line.

The Robinson-Trautman metrics and the Kerr and charged Kerr metrics with
their properties are explicit examples of the construction given here.
\end{abstract}

\section{\textbf{Introduction}}

When discussing or studying the general properties of a set of differential
equations, it is usually best to try to adopt a coordinate system to the
symmetries of the manifold in question. For example general properties of
Maxwell's equations are best studied using standard Minkowski coordinates.
If however one is looking for solutions with specific properties, symmetries
for example, it is clearly best to adopt coordinates closest to those of the
solutions. Spherically symmetric solutions are most easily studied in
spherical coordinates. In general relativity, when studying general
properties of asymptotically flat space-times\cite{TN} the most used
coordinate/tetrad system, in the neighborhood of future null infinity, $%
\frak{I}^{+}$, are the Bondi coordinate/tetrad system\cite{NP} which is
associated with the asymptotic symmetry group, i.e., with the
Bondi-Metzner-Sachs group\cite{P3}. On the other hand, if \textit{particular}
asymptotically flat solutions are to be found, special non-Bondi coordinates
and/or tetrads are often used. One sees this, for example in the study of
the Robinson-Trautman metrics or the twisting type II metrics where both
different coordinates and different tetrad systems are often used.

In this work we will show that for every specific asymptotically flat
space-time there is a non-Bondi coordinate/tetrad system made-to-order for
that solution that simplifies the discussion of the solution and brings out
its properties more clearly than does the Bondi system. Often one can use
the Bondi coordinates with the non-Bondi tetrad, though for certain
situations, discussed later, both the non-Bondi coordinates and tetrads are
preferred

To approach this issue, in Sec.II, we give a brief review and survey of the
relevant properties of $\frak{I}^{+}$ that are needed here. This includes
the possible use of two different types of coordinates, Bondi and NU
coordinates\cite{TN}, and two different types of null tetrad systems, (1)
those based on a surface forming [twist-free] null vector field near
`infinity' and (2) those based on a twisting null vector field. In Sec.III,
we discuss what are the different relevant functions that are defined and
used on $\frak{I}^{+},$ (for any given space-time), that are used in the
different $\frak{I}^{+}$ coordinate/tetrad systems. It will be seen that
depending on the different choices of coordinate/tetrad system, the
descriptions of the \textit{same physical situation} can be very different
and that large simplifications are available by the appropriate choices. In
Sec.IV we will display the Bianchi Identities for the different functions in
the different coordinate and tetrad systems; all quite different from each
other. In Sec.V we discuss how one tries to extract physical quantities from
the variables on $\frak{I}^{+}.$

The major development reported in this work is given in Sec.VI. We show that
for \textit{any asymptotically flat space-time} there is a coordinate/tetrad
(non-Bondi) system that can be adapted to the specific space-time that
allows a considerable simplification in the relevant equations. Furthermore
it allows us to give a definition of angular momentum (orbital and spin)
that is geometric and hence independent of the choice of coordinates and
thus is supertranslation invariant. The point of view that is adopted is as
follows: an arbitrary asymptotically flat space-time is first completely
described in a Bondi coordinate/tetrad frame by the freely chosen \textit{\
characteristic data}, the asymptotic shear, $\sigma ,$ and some (irrelevant
for us) initial data\textit{. \{We have tacitly assumed throughout that the
space-time can be described or approximated by real analytic functions.\}}
Keeping the Bondi coordinates, we now change the tetrad, in a specific way:
we go from the Bondi tetrad which has non-vanishing asymptotic shear (our
free data $\sigma $) to a new tetrad that has (in general) \textit{twist}
but is \textit{\ asymptotically shear free}. All the information that was in
the shear is thus put into the geometry of the \textit{twisting} \textit{%
shear free }tetrad. The relevant function to describe the twisting tetrad is
a complex (stereographic) angle at each point of $\frak{I}^{+}$ that is
denoted by $L[\frak{I}^{+}].$ This process is reversible. Given $L,$ the $%
\sigma $ can be reconstructed. The new feature is that the \textit{\
shear-free twisting} congruence, given by $L,$ is \textit{not} uniquely
determined by the original shear. The remaining freedom in $L$ is given by
four complex functions of a single complex variable which can be thought of,
locally, as a (complex) curve, $\xi ^{a}(\tau ),$ in \textbf{C}$^{4}.$
[Specifically and more surprising, it turns out to be a curve in $H$ -Space%
\cite{H-space1,H-space2}.] At this point it is a completely arbitrary curve,
i.e., any choice of the curve is related to the same shear. We show that
there is a natural condition that can be imposed on the curve that makes it
unique (up to initial data) that leads us to \textit{define} the spin
angular momentum and the center of mass from the real and imaginary parts of
the complex curve. These definitions are geometrically invariant under Bondi
supertranslation and reduce to the standard definitions for the Kerr metric.
When the asymptotic shear is pure `electric type' the spin vanishes, i.e.,
the spin arises from the `magnetic' part of the shear. Unfortunately to do
this construction exactly is quite difficult and one must resort to
approximations to obtain approximate explicit results. Fortunately there is
a straightforward approximation method which is summarized in an appendix.

In Sec.VII we give and discuss several examples. They include the
Robinson-Trautman metrics, the type II twisting metrics and space-times
obtained from the class of shears referred to as the Tod-Sparling\cite{TS}
shears.

Though everything that is said and done here \textit{could} have been stated
in the language of the conformal compactification of space-time, we will
retain the physical space-time description. From this point of view, the
null boundary is reached by letting our null geodesics go to future
infinity, i.e., by letting the geodesic affine parameters, $r$, go to
infinity.

We assume that the reader is familiar with spin-coefficient formalism and
the associated spin-$s$ harmonics though for completeness we have included
in Appendix C several of the useful relations.

In addition for comparison with other works, we mention the awkward
convention adopted here for historic reasons\cite{TN} and from frequent
usage, that our radial coordinate $r$ differs from the more conventional
choice of $r_{c},$ i.e., by $r=\sqrt{2}r_{c}$ and our $u_{B}=u_{c}/\sqrt{2}.$
This leads, in the Schwarzschild case, to the unconventional form 
\begin{eqnarray}
g_{00}du_{B}^{2} &=&2(1-\frac{2\sqrt{2}GM}{c^{2}r})du_{B}^{2}
\label{convention} \\
\psi _{2}^{0} &=&-\frac{2\sqrt{2}GM}{c^{2}}.  \nonumber
\end{eqnarray}

\section{\textbf{Review of Scri}$^{+}$\textbf{: the Future Null Boundary of
Space-Time}}

\subsection{Scri coordinates}

Future null infinity, often referred to as $\frak{I}^{+},$ is roughly
speaking the set of endpoints of all future directed null geodesics\cite{P2}%
. It is usually given the structure of $S^{2}xR,$ a line bundle over the
sphere. This leads naturally to the global stereographic coordinates, ($%
\zeta ,\overline{\zeta },),$ for the $S^{2}$ and the $u$ labeling the cross
sections or global slices of the bundle. From the point of view of the
space-time, $\frak{I}^{+}$ is a null surface and the null generators are
given by ($\zeta ,\overline{\zeta },)$ $=constant$.

There is a canonical slicing of $\frak{I}^{+}$, $u_{B}=constant$, referred
to as the Bondi slicing, that it is defined from an asymptotic symmetry
inherited from the interior, called the Bondi-Metzner-Sachs group. Any one
Bondi slicing is related to any other by the \textit{supertranslation}
freedom;

\begin{equation}
\widetilde{u}_{B}=u_{B}+\alpha (\zeta ,\overline{\zeta }).  \label{ST}
\end{equation}
with $\alpha (\zeta $,$\overline{\zeta })$ an arbitrary regular function on $%
S^{2}.$

\quad All other arbitrary slicings, ($u=\tau ),$ often called NU slicings,
are given by 
\begin{equation}
\quad u_{B}=G(\tau ,\zeta ,\overline{\zeta }).  \label{NUT}
\end{equation}

\subsection{Null Tetrads on Scri}

In addition to the choice of coordinate systems there is the freedom to
chose the asymptotic null tetrad system, ($l^{a},m^{a},\overline{m}
^{a},n^{a} $). Since $\frak{I}^{+}$ is a fixed null surface with its own
generators, with null tangent vectors, [say $n^{a}],$ it is natural to keep
it fixed. The remaining tetrad freedom lies, essentially, in the choice of
the other null leg, $l^{a}.$ Most often $l^{a}$ is chosen to be orthogonal
to the Bondi, $u_{B}=constant,$ slices. In this case we will refer to a
Bondi coordinate/tetrad system. Relative to such a system any other tetrad
set, ($l^{*a},m^{*a},n^{*a}$), is given by a null rotation\cite{Aronson}
about $n^{a},$ i.e., 
\begin{eqnarray}
l^{*a} &=&l^{a}+b\overline{m}^{a}+\overline{b}m^{a}+b\overline{b}n^{a},
\label{NullRot} \\
m^{*a} &=&m^{a}+bn^{a},  \nonumber \\
n^{*a} &=&n^{a},  \nonumber \\
b &=&-L/r+O(r^{-2}).  \nonumber
\end{eqnarray}

The complex function $L(u_{B},\zeta ,\overline{\zeta },)$, the
(stereographic) angle between the null vectors, $l^{*a}$ and $l^{a},$ is at
this moment completely arbitrary but later will be dynamically determined.
Depending on how $L(u_{B},\zeta ,\overline{\zeta },)$ is chosen $l^{*a}$
might or might not be surface forming. We will abuse the language/notation
and refer to the tetrad systems associated with $l^{*a}$ as the \textit{%
twisting-type} tetrads as distinct from a Bondi tetrad even when $l^{*a}$ is
surface forming.

The function $L(u_{B},\zeta ,\overline{\zeta },)$ and its choice will later 
\textit{play the pivotal role} in this work. It will be seen that the vector
field $l^{*a}$ can be constructed by an appropriately chosen $L(u,\zeta ,%
\overline{\zeta },)$ so that it is asymptotically \textit{shear-free}.

\section{\textbf{Quantities Defined on }$\frak{I}^{+}$\textbf{\ for Any
Given Interior Space-Time}}

Using the spin-coefficient notation for the five complex components of the
asymptotic Weyl tensor, we have, from the peeling theorem\cite{TN,NP}, that 
\begin{eqnarray*}
\psi _{0} &=&\frac{\psi _{0}^{0}}{r^{5}}+0(r^{-6}) \\
\psi _{1} &=&\frac{\psi _{1}^{0}}{r^{4}}+0(r^{-5}) \\
\psi _{2} &=&\frac{\psi _{2}^{0}}{r^{3}}+0(r^{-4}) \\
\psi _{3} &=&\frac{\psi _{3}^{0}}{r^{2}}+0(r^{-3}) \\
\psi _{4} &=&\frac{\psi _{4}^{0}}{r}+0(r^{-2})
\end{eqnarray*}
where the five, $\psi _{4}^{0},\psi _{3}^{0},\psi _{2}^{0},\psi
_{1}^{0},\psi _{0}^{0},$ are functions defined on $\frak{I}^{+}.$ For a
given space-time their explicit expressions depend on the choices of \textit{%
\ both} coordinates and tetrads. Their evolution is determined by the
asymptotic Bianchi identities and the \textit{type of characteristic data}
that is given.

We adopt the following notation: expression written in the Bondi
coordinate/tetrad system will appear without a star,`$^{*}$', e.g., $\psi
_{2}^{0},$ while for Bondi coordinates with a twisting tetrad a `$^{*}$'
will be used, e.g., $\psi _{2}^{*0}$. In the case of NU coordinates and
tetrad a double star will be used, e.g. $\psi _{2}^{**0},$etc.

Using Bondi coordinates and tetrad, the free characteristic data is given
only by the (complex) asymptotic shear,

\[
\sigma =\sigma (u_{B},\zeta ,\overline{\zeta }), 
\]
while in Bondi coordinates, but with a twisting-type tetrad the free
functions are 
\[
\sigma ^{*}=\sigma ^{*}(u_{B},\zeta ,\overline{\zeta })\text{ \& }%
L(u_{B},\zeta ,\overline{\zeta }) 
\]
with $\sigma ^{*}(u_{B},\zeta ,\overline{\zeta })$ and $L(u_{B},\zeta ,%
\overline{\zeta })$ carrying the same (redundant) information as did the
Bondi $\sigma (u_{B},\zeta ,\overline{\zeta }).$

In the case of NU coordinates and tetrad, the free data is given by

\[
\sigma ^{**}=\sigma ^{**}(\tau ,\zeta ,\overline{\zeta })\text{ \& }V(\tau
,\zeta ,\overline{\zeta })\text{ } 
\]
with 
\[
V(\tau ,\zeta ,\overline{\zeta })\equiv du_{B}/d\tau =\partial _{\tau
}G(\tau ,\zeta ,\overline{\zeta }). 
\]

The important point is that in both cases, the Bondi coordinates with
twisting tetrad and the NU coordinates and tetrad, all the information that
was in $\sigma (u,\zeta ,\overline{\zeta })$ is transferred to the new
variables, ($\sigma ^{*}$\& $L)$ or ($\sigma ^{**}$\& $V)$. Later we will
show that \textit{all the information} can be shifted into an appropriately
chosen $L(u_{B},\zeta ,\overline{\zeta })$ with a vanishing $\sigma ^{*}$
and that in certain special cases (a pure `electric' type $\sigma $) \textit{%
all the information} can be shifted into the $V(\tau ,\zeta ,\overline{\zeta 
})$ with vanishing $\sigma ^{**}.$

The relationship between the ($\psi _{4}^{0},\psi _{3}^{0},\psi
_{2}^{0},\psi _{1}^{0},\psi _{0}^{0})$ given in a Bondi tetrad and the ($%
\psi _{4}^{*0},\psi _{3}^{*0},\psi _{2}^{*0},\psi _{1}^{*0},\psi _{0}^{*0})$
of a twisting tetrad is 
\begin{eqnarray}
\psi _{0}^{0} &=&\psi _{0}^{*0}+4L\psi _{1}^{*0}+6L^{2}\psi
_{2}^{*0}+4L^{3}\psi _{3}^{*0}+L^{4}\psi _{4}^{*0}  \label{BondiToTwist} \\
\psi _{1}^{0} &=&\psi _{1}^{*0}+3L\psi _{2}^{*0}+3L^{2}\psi
_{3}^{*0}+L^{3}\psi _{4}^{*0}  \label{BtoT2} \\
\psi _{2}^{0} &=&\psi _{2}^{*0}+2L\psi _{3}^{*0}+L^{2}\psi _{4}^{*0}
\label{BtoT3} \\
\psi _{3}^{0} &=&\psi _{3}^{*0}+L\psi _{4}^{*0}  \label{BtoT4} \\
\psi _{4}^{0} &=&\psi _{4}^{*0}.  \label{BtoT5}
\end{eqnarray}

This is used, in the next section, going from the Bondi version of the
Bianchi identities to the twisting version. To go to the conventional NU
tetrad a further scaling is needed\cite{TN}.

\section{The Asymptotic Bianchi Identities}

In each of the different coordinate/tetrad cases there are relations and
differential equations relating the functions defined on $\frak{I}^{+}$.
Specifically, the $\psi _{3}^{0}$ \& $\psi _{4}^{0}$, in each of the cases,
are given explicitly in terms of the functions, $\sigma $, $L$ or $V.$ The
dynamics lies in the differential equations for the remaining $\psi ^{\prime
}s,.i.e.,$ ($\psi _{0}^{0\,},\psi _{1}^{0\,},\psi _{2}^{0\,}$).

Explicitly, in a Bondi coordinate and tetrad system, they are:

\begin{eqnarray}
\psi _{0}^{0\,\cdot } &=&-\text{ $\frak{d}$}\psi _{1}^{0}+3\sigma \psi
_{2}^{0}  \label{Bondi1} \\
\psi _{1}^{0\,\cdot } &=&-\text{ $\frak{d}$}\psi _{2}^{0}+2\sigma \psi
_{3}^{0}  \label{Bondi2} \\
\psi _{2}^{0\,\cdot } &=&-\text{ $\frak{d}$}\psi _{3}^{0}+\sigma \psi
_{4}^{0}  \label{Bondi3} \\
\psi _{3}^{0} &=&\text{ $\frak{d}$}\overline{\sigma }^{\cdot }
\label{Bondi4} \\
\psi _{4}^{0} &=&-\overline{\sigma }^{\cdot \cdot }  \label{Bondi5} \\
\psi _{2}^{0\,}-\overline{\psi }_{2}^{0\,} &=&\overline{\frak{d}}^{2}\sigma -%
\text{ $\frak{d}$}^{2}\overline{\sigma }+\overline{\sigma }\sigma ^{\cdot %
}-\sigma \overline{\sigma }^{\cdot },  \label{Bondi6} \\
&&  \nonumber \\
\text{ }\Psi  &=&\psi _{2}^{0\,}+\text{ $\frak{d}$}^{2}\overline{\sigma }%
+\sigma \overline{\sigma }^{\cdot }  \label{Bondi8} \\
\Psi -\overline{\Psi } &=&0  \label{Bondi9}
\end{eqnarray}

For later use we display the same set of equations in both a
Bondi-coordinate/twisting Tetrad system and in NU-coordinate/tetrad system
but where the simplification 
\[
\sigma ^{*}=0\text{ and }\sigma ^{**}=0 
\]
was used. The justification for this simplification is given in the next
section.

\begin{remark}
For a Bondi Coordinate/Twisting Tetrad system, with $\sigma ^{*}=0,$ we
have, from Eq.(\ref{BondiToTwist}), that 
\begin{eqnarray}
\psi _{0}^{*0\,\cdot } &=&-\text{ $\frak{d}$}\psi _{1}^{*0}-L\psi _{1}^{*0\,%
\cdot }-4\dot{L}\psi _{1}^{*0}  \label{twisting1} \\
\psi _{1}^{*0\,\cdot } &=&-\text{ $\frak{d}$}\psi _{2}^{*0}-L\psi _{2}^{*0\,%
\cdot }-3\dot{L}\psi _{2}^{*0}  \label{twisting2} \\
\psi _{2}^{*0\,\cdot } &=&-\text{ $\frak{d}$}\psi _{3}^{*0}-L\,\psi
_{3}^{*0\,\cdot }-2\dot{L}\psi _{3}^{*0}  \label{twisting3} \\
\psi _{3}^{*0} &=&\text{ $\frak{d}$}\overline{\sigma }^{\cdot }+L\overline{%
\sigma }^{\cdot \cdot }  \label{twisting4} \\
\psi _{4}^{*0} &=&-\overline{\sigma }^{\cdot \cdot }  \label{twisting5} \\
\sigma  &\equiv &\text{ $\frak{d}$}L+\frac{1}{2}(L^{2})^{\cdot }
\label{twisting6} \\
\Psi  &=&\psi _{2}^{*0}+2L\text{ $\frak{d}$}\overline{\sigma }^{\cdot }+L^{2}%
\overline{\sigma }^{\cdot \cdot }+\text{ $\frak{d}$}^{2}\overline{\sigma }%
+\sigma \overline{\sigma }^{\cdot }  \label{twisting7} \\
\Psi -\overline{\Psi } &=&0  \label{twisting8*}
\end{eqnarray}
\end{remark}

\begin{remark}
Eqs. (\ref{Bondi8}) and (\ref{twisting7}) are identical even though they
appear different. This can be seen by using Eq.(\ref{BtoT3}).
\end{remark}

For a NU coordinate/tetrad system, with $\sigma ^{**}=0,$ after rescaling
the $\psi $'$s$ with appropriate powers of $V(\tau ,\zeta ,\overline{\zeta })
$, 
\begin{eqnarray}
\psi _{0}^{**0\,\prime } &=&-\text{ $\frak{d}$}\psi _{1}^{**0}+3\frac{%
V^{\prime }}{V}\psi _{0}^{**0}  \label{NU1} \\
\psi _{1}^{**0\,\prime } &=&-\text{ $\frak{d}$}\psi _{2}^{**0}+3\frac{%
V^{\prime }}{V}\psi _{1}^{**0}  \label{NU2} \\
\psi _{2}^{**0\,\prime } &=&-\text{ $\frak{d}$}\psi _{3}^{**0}+3\frac{%
V^{\prime }}{V}\psi _{2}^{**0}  \label{NU3} \\
\psi _{3}^{**0} &=&\overline{\frak{d}}^{2}\text{ $\frak{d}$}\log V
\label{NU4} \\
\psi _{4}^{**0} &=&-\overline{\frak{d}}^{2}\frac{V^{\prime }}{V}  \label{NU5}
\\
\psi _{2}^{**0\,}-\overline{\psi }_{2}^{**0\,} &=&0  \label{NU6}
\end{eqnarray}

In the first two cases, the dot indicates $u_{B}$-derivatives, while in the
NU case the prime means the $\tau $ derivative.

We want to emphasize that we have not lost any generality when going to the
Bondi/Twisting tetrad but the NU equations \textit{are valid only when the
original Bondi shear} $\sigma (u_{B},\zeta ,\overline{\zeta })$ \textit{was
pure `electric'}. In other words all the information in an `electric' shear
can be put into the real $V(u_{B},\zeta ,\widetilde{\zeta }$). Details will
be given in Sec.VI.

\section{\textbf{The Physical Quantities}}

\quad Most often one extracts or attempts to extract the physical quantities
by trying to identify them in the midst of this set of functions at $\frak{I}%
^{+}$ by integrals over the $\psi ^{\prime }$s and complicated combinations
of the $\sigma .$ In general this a difficult and frequently ambiguous task
- largely because of the Bondi super-translation freedom. Only in the case
of the identification of four-momentum has it been done successfully and
unambiguously. In Bondi coordinates and tetrad, one defines the mass aspect $%
\Psi (u_{B},\zeta ,\overline{\zeta }),$ which is real from Eqs.(\ref{Bondi8}%
) and (\ref{Bondi9}), by 
\begin{equation}
\Psi =\psi _{2}^{0}+\text{ $\frak{d}$}^{2}\overline{\sigma }+\sigma 
\overline{\sigma }^{\cdot }.  \label{Maspect}
\end{equation}

The four-momentum is extracted from the $l=0$ \& $1$ coefficients of the
spherical harmonics. When Eq.(\ref{Bondi3}) is expressed in terms of $\Psi $
it takes the very simple form 
\begin{equation}
\Psi ^{\cdot }=\sigma ^{\cdot }\overline{\sigma }^{\cdot }
\label{BondiMassLoss}
\end{equation}
which, when integrated over the sphere, becomes the Bondi mass loss
equation. These results can be transformed and reexpressed via the twisting
tetrads or in the NU coordinate/tetrad system. Explicitly, Eq.(\ref
{twisting3}) becomes identical to Eq.(\ref{BondiMassLoss}) except that now 
\[
\sigma ^{\cdot }\equiv \text{ $\frak{d}$}L^{\cdot }+\frac{1}{2}(L^{2})^{%
\cdot \cdot }.
\]

Unfortunately there is no agreed upon asymptotic definition of orbital or
spin angular momentum.

In the following section we will address this issue from a totally different
point of view. Up to now the physical quantities were assumed to be
contained in the asymptotic Weyl tensor and shear and constructed by a
kinematic process.

We propose that rather than the kinematic process, there is a dynamic
process involving the known functions on $\frak{I}^{+}$ and only after it is
solved can the orbital or spin angular momentum be found.

\section{A \textbf{New Idea - Dynamics on }$\frak{I}^{+}$}

We first will show that there are special choices of $L(u_{B},\zeta ,%
\overline{\zeta })$ leading to the twisting tetrads for which the shear, $%
\sigma ^{*}=0.$\ To find this family of functions $L(u_{B},\zeta ,\overline{%
\zeta }),$ a differential equation must be solved. The solution, however, is
not unique. The freedom in the solution is four arbitrary complex functions
of a single complex variable, i.e., it is an arbitrary complex curve in a
four-dimensional parameter space. [The parameter space is the well-studied $%
H $-space.] Note that all the information in the Bondi characteristic data, $%
\sigma (u_{B},\zeta ,\overline{\zeta }),$ will have been shifted to the $%
L(u_{B},\zeta ,\overline{\zeta }).$ Given an $L(u_{B},\zeta ,\overline{\zeta 
})$ with \textit{any one of these curves,} we could go backwards to recover
the original Bondi $\sigma (u_{B},\zeta ,\overline{\zeta }).$ The dynamics
lies in the unique determination of this curve from other considerations
which are describe later. The real part of this curve determines the center
of mass (or orbital momentum) while the imaginary part yields the
spin-angular momentum\cite{PM,N,LN}. These quantities are invariant under
supertranslations.

We begin by observing that if we start with a Bondi coordinate/tetrad system
with a given shear $\sigma (u_{B},\zeta ,\overline{\zeta }),$ then [after an
unpleasant calculation\cite{Aronson}] the shear of the new null vector $%
l^{*a}$ after the null rotation, Eqs.(\ref{NullRot}), is related to the old
one by 
\[
\sigma ^{*}(u_{B}^{*},\zeta ,\overline{\zeta })=\sigma (u_{B},\zeta ,%
\overline{\zeta })-\text{ $\frak{d}$}L-LL^{\cdot }.
\]

The function $L(u,\zeta ,\overline{\zeta })$ is then chosen so that the new
shear vanishes, i.e., it must satisfy 
\begin{equation}
\text{ $\frak{d}$}L+LL^{\cdot }=\sigma (u_{B},\zeta ,\overline{\zeta }).
\label{ShearFree}
\end{equation}

\begin{remark}
The special case, $\frak{d}L+LL^{\cdot }=0,$ played an important early role
in the development of twistor theory, leading immediately to the
relationship in flat space between shear-free null geodesic congruences and
twistor theory. It leads us to conjecture that Eq.(\ref{ShearFree}) might
play a role in some asymptotic form of twistor theory.
\end{remark}

Though Eq.(\ref{ShearFree}) is non-linear with an arbitrary right-side and
appears quite formidable, considerable understanding of it can be found by
the following procedure:

Assume the existence of a \textit{complex} function 
\begin{equation}
\tau =T(u_{B},\zeta ,\overline{\zeta })  \label{tau}
\end{equation}
that is invertible in the sense that it can, in principle, be written as 
\begin{equation}
u_{B}=X(\tau ,\zeta ,\overline{\zeta }).  \label{X}
\end{equation}
Then writing 
\begin{equation}
L=-\frac{\text{ $\frak{d}$}T}{T^{\cdot }},  \label{L1}
\end{equation}
and using the implicit derivatives of Eq.(\ref{X}), 
\begin{eqnarray}
1 &=&X,_{\tau }T^{\,\cdot }  \label{Implicit} \\
0 &=&\text{ $\frak{d}$}_{(\tau )}X+X^{\prime }\text{ $\frak{d}$}T,  \nonumber
\end{eqnarray}
we obtain 
\begin{equation}
L=\text{ $\frak{d}$}_{(\tau )}X.  \label{L2}
\end{equation}
Prime means the $\tau $ derivative and  $\frak{d}_{(\tau )}$ means  $\frak{d}
$\ with $\tau $ held constant. Finally, with this implicit view of $X$ and $T
$, Eq.(\ref{ShearFree}) becomes 
\begin{mathletters}
\begin{equation}
\text{ $\frak{d}$}_{(\tau )}^{2}X=\sigma (X,\zeta ,\overline{\zeta }).
\label{GoodCut}
\end{equation}

\begin{remark}
Note that this equation, derived here from a different perspective, is
identical to the so-called \textit{good cut equation\cite{H-space1,H-space2}.%
} In general it has a four complex dimensional solution space that has been
dubbed $H$-space. $H$-space possesses a complex metric which is Ricci-flat
with a self-dual Weyl tensor. It formed the basis of Penrose's asymptotic
twistor space\cite{PM}. In its original form it had its origin in the search
for complex null surfaces that were asymptotically shear-free. Here we are
looking for \textit{real} null geodesic congruences that are, in general,
not surface forming, i.e., have twist, but are asymptotically shear-free\cite
{KN2}.
\end{remark}

\end{mathletters}
\begin{remark}
Note that using the reparametrization 
\begin{equation}
\tau \Rightarrow \tau ^{*}=F(\tau )=T^{*}(u_{B},\zeta ,\overline{\zeta })
\label{repar}
\end{equation}

leaves Eq.(\ref{L1}) and the entire construction invariant. This fact will
be used later for certain simplifications.
\end{remark}

\begin{remark}
We mention that if the shear in Eq.(\ref{GoodCut}), is pure electric
[i.e.,\thinspace $\sigma (u_{B},\zeta ,\overline{\zeta })$ $=$ $\frak{d}%
^{2}S(u_{B},\zeta ,\overline{\zeta })$ with $S(u_{B},\zeta ,\overline{\zeta }%
)$ a \textit{real function}] then the associated $H$-space is flat and has a
real four-dimensional subspace that can be identified with Minkowski space.
We return to this issue later.
\end{remark}

As we just mentioned, the solutions to Eq.(\ref{GoodCut}) depend on four
complex parameters, say $z^{a},$ and can be summarized by

\begin{equation}
u_{B}=X(z^{a},\zeta ,\overline{\zeta }).  \label{H}
\end{equation}
However, since we are interested in solutions of the form, Eq.(\ref{X}),
i.e., 
\begin{equation}
u_{B}=X(\tau ,\zeta ,\overline{\zeta }),  \label{X*}
\end{equation}
all we must do is chose in, $H$-space, an arbitrary complex world-line, $%
z^{a}=\xi ^{a}(\tau ),$ and substitute it into Eq.(\ref{H}) to obtain the
form (\ref{X*}). Furthermore, it can be seen that the solution, (\ref{X*}),
can be written as 
\[
u_{B}=X(\tau ,\zeta ,\overline{\zeta })=\xi ^{a}(\tau )\widehat{l}_{a}(\zeta
,\overline{\zeta })+X_{(l\geqslant 2)}(\tau ,\zeta ,\overline{\zeta })
\]
with $X_{(l\geqslant 2)}$ containing spherical harmonics, $l\geqslant 2$ and 
\begin{equation}
\widehat{l}_{a}(\zeta ,\overline{\zeta })=\frac{\sqrt{2}}{2}(1,-\frac{\zeta +%
\overline{\zeta }}{1+\zeta \overline{\zeta }},i\frac{\zeta -\overline{\zeta }%
}{1+\zeta \overline{\zeta }},\frac{1-\zeta \overline{\zeta }}{1+\zeta 
\overline{\zeta }})  \label{X1.2}
\end{equation}
which has only the $l=0$ \& $1,$ spherical harmonics. This result follows
from the fact that  $\frak{d}_{(\tau )}^{2}$ in the good-cut equation
annihilates the $l=0$ \& $1$ terms. The $\xi ^{a}(\tau )$ does get fed back
into the higher harmonics via the $\sigma (X,\zeta ,\overline{\zeta }).$

Our solution to Eq.(\ref{ShearFree}) can now be expressed implicitly by 
\begin{eqnarray}
L(u_{B},\zeta ,\overline{\zeta }) &=&\text{ $\frak{d}$}_{(\tau )}X=\xi
^{a}(\tau )\widehat{m}_{a}(\zeta ,\overline{\zeta })+\text{ $\frak{d}$}%
_{(\tau )}X_{(l\geqslant 2)}(\tau ,\zeta ,\overline{\zeta })  \label{L(u)} \\
u_{B} &=&X=\xi ^{a}(\tau )\widehat{l}_{a}(\zeta ,\overline{\zeta }%
)+X_{(l\geqslant 2)}(\tau ,\zeta ,\overline{\zeta }).  \label{u2}
\end{eqnarray}

\begin{remark}
We emphasize that the complex world-line, $\xi ^{a}(\tau ),$ is not in
physical space but is in the parameter space, $H$-space. At this point we
are not suggesting that there is anything profound about this observation.
It is simply there and whatever meaning it might have is obscure. We however
remark that this observation can be extended to the Einstein-Maxwell
equations where in general the Maxwell field will have its own complex
world-line. If the two independent complex world lines coincide, [from
preliminary results and special cases] one obtains the Dirac value of the
gyromagnetic ratio\cite{TedDublin}.
\end{remark}

\begin{remark}
To invert the later equation, (\ref{u2}), to obtain $\tau =T(u_{B},\zeta ,%
\overline{\zeta }),$ is, in general, virtually impossible. There however is
a relatively easy method, by iteration, to get approximate inversions to any
accuracy. \{See Appendix A\}
\end{remark}

\begin{remark}
We have tacitly assumed that the world line is complex analytic in the 
\textit{complex parameter} $\tau .$

If we expand $\xi ^{a}(\tau )$ in a Taylor series and regroup the terms with
real coefficients and those with imaginary coefficients separately, we can
write 
\[
\xi ^{a}(\tau )=\xi _{R}^{a}(\tau )+i\xi _{I}^{a}(\tau ).
\]
This decomposition becomes important later.
\end{remark}

\begin{remark}
The asymptotic twist, $\Sigma (u_{B},\zeta ,\overline{\zeta }),$ \{a measure
of how far the vector field $l^{*a}$ is from being surface forming\} is
defined by 
\[
2i\Sigma =\text{ $\frak{d}$}\overline{L}+L\overline{L}^{\cdot }-\text{ }%
\overline{\frak{d}}L-\overline{L}L^{\cdot }.
\]
If $L$ is obtained from a shear that is pure `electric', then the $\xi
^{a}(\tau )$ can be chosen so that the asymptotic twist vanishes.
\end{remark}

It is from the fact that we can chose $L(u,\zeta ,\overline{\zeta })\ $as a
solution of Eq.(\ref{ShearFree}), leading to

$\sigma ^{*}=\sigma ^{**}=0,$ that is the justification for the form of Eqs.(%
\ref{twisting1})-(\ref{twisting8*}).

\subsection{Determining the Complex Curve}

The argument and method for the unique determination of the complex curve
(and eventually the definitions of spin and center of mass motion) are not
along conventional lines of thought. To try to clarify it, it is worthwhile
to take a brief detour.

It is well-known that for a static charge distribution, the electric dipole
moment is given by the total charge times the center of charge position
given in the `static' Lorentz frame. In the case of a dynamic charge it is
more difficult since the center of charge depends on the choice of Lorentz
frame. Nevertheless by having, in any one frame, the center of charge at the
coordinate origin, the electric dipole moment will vanish. In the case of a
single charged particle moving on an arbitrary real world line, (the
Lienard-Wiechert Maxwell field) the electric dipole moment vanishes when the
coordinate origin follows the particle's motion or equivalently, the dipole
is given by the particles displacement from the coordinate origin times the
charge. The asymptotically defined dipole moment also vanishes when it is
calculated (extracted) from the $l=1$ part of the coefficient of the $r^{-3}$
term of $\phi _{0}=F_{ab}l^{a}m^{b},$ where $l^{a}$ is a tangent vector of
the light-cones that are attached to the particles world-line. Analogously,
it was shown\cite{ShearFreeMax} that the magnetic dipole moment of a charged
particle can be `viewed' as arising from a charge moving in complex
Minkowski space along a complex world-line. It is given by the charge times
the imaginary displacement. \{We emphasize again that the complex
world-lines are a bookkeeping device and no claim is made that particles are
really moving in complex space. The \textit{real} effect of the complex
world-line picture is to create a twisting real null congruence.\} If the
light-cone from the complex world-line is followed to $\frak{I}^{+},$
(equivalent to introducing an appropriate asymptotic twist for the null
vector $l^{a}$ that is used in $\phi _{0}=F_{ab}l^{a}m^{b}$) then both the
asymptotic electric and magnetic dipoles vanish. This argument is given in
considerable detail in the paper\cite{ShearFreeMax,KN}.

The idea is to generalize this construction to GR. In linearized GR and in
the exact case of the Kerr and charged Kerr metrics, the mass-dipole moment
and the spin sit in the real and imaginary parts of the $l=1$ harmonic of
the coefficient of the $r^{-4}$ part of the Weyl component $\psi _{1}^{0\,}$
in a Bondi coordinate/tetrad system. This observation is extended to all
asymptotically flat vacuum solutions by going to an asymptotically twisting
tetrad, with an appropriately chosen $H$-space complex curve [see below],
such that the new Weyl component, $\psi _{1}^{*0\,},$ has a \textit{vanishing%
} $l=1$ harmonic in the coefficient of its $r^{-4}$ part. We refer to this
curve as the intrinsic complex center of charge world-line and, rough
speaking, the center of mass and the spin angular momentum will be
identified by its real and imaginary parts. Effectively, by chosing the
world-line so that the $l=1$ harmonic vanishes we have shifted the `origin'
to the complex world-line.

To explicitly apply this idea to our discussion of twisting null
congruences, we go to the set of equations, Eq.(\ref{twisting2}) and (\ref
{twisting3}), namely 
\begin{eqnarray}
\psi _{1}^{*0\,\cdot } &=&-\text{ $\frak{d}$}\psi _{2}^{*0}-L\psi _{2}^{*0\,%
\cdot }-3\dot{L}\psi _{2}^{*0}  \label{twisting2*} \\
\psi _{2}^{*0\,\cdot } &=&-\text{ $\frak{d}$}\psi _{3}^{*0}-L\,\psi
_{3}^{*0\,\cdot }-2\dot{L}\psi _{3}^{*0}  \label{twisting3*}
\end{eqnarray}
and require that, in (\ref{twisting2*}), that the $l=1$ part of $\psi
_{1}^{*0\,}$ should vanish. This leads (via approximations) to a
differential equation for the complex center-of-mass world-line. It is
driven by the original Bondi shear. To see this in detail, first we recall
that (\ref{twisting3*}) can be rewritten as

\begin{eqnarray}
\Psi ^{\cdot } &=&\sigma ^{\cdot }\overline{\sigma }^{\cdot }  \label{1} \\
\sigma ^{\cdot } &\equiv &\text{ $\frak{d}$}L^{\cdot }+\frac{1}{2}(L^{2})^{%
\cdot \cdot }  \label{2} \\
\psi _{2}^{*0} &=&\Psi -2L\text{ $\frak{d}$}\overline{\sigma }^{\cdot }-L^{2}%
\overline{\sigma }^{\cdot \cdot }-\text{ $\frak{d}$}^{2}\overline{\sigma }%
-\sigma \overline{\sigma }^{\cdot }.  \label{3}
\end{eqnarray}

If we use the result [see Appendix B] that the Bondi four-momentum, $P_{a},$
can be extracted from $\Psi $ by

\begin{equation}
P_{a}(u_{B})\equiv (M,-P^{i})=-\frac{c^{2}}{8\pi G}\int \Psi \widehat{l}%
_{a}dS  \label{P}
\end{equation}
we can invert and see that

\begin{equation}
\Psi (u_{B},\zeta ,\overline{\zeta })\equiv \chi -\chi ^{i}\widehat{c}%
_{i}+....=-2\frac{G}{c^{2}}(\sqrt{2}M-3P^{i}\widehat{c}_{i})+\Psi _{l\geq
2}...  \label{PSI.II}
\end{equation}
with

\begin{eqnarray}
\widehat{l}_{b}(\zeta ,\overline{\zeta }) &=&\frac{\sqrt{2}}{2}%
(1,C_{i}(\zeta ,\overline{\zeta })),  \label{etc1} \\
\widehat{c}_{b} &\equiv &\widehat{l}_{b}-\widehat{n}_{b}=\sqrt{2}%
(0,C_{i}(\zeta ,\overline{\zeta }))  \label{etc*} \\
C_{a} &=&\frac{\sqrt{2}}{2}\widehat{c}_{a},  \label{etc**}
\end{eqnarray}
$C_{i}$ a unit radial vector and $\Psi _{l\geq 2}$ containing only harmonics 
$l\geq 2.$

The time evolution of $P_{a}(u)$ is found by integration Eq.(\ref{1}), i.e.,
from $\Psi =\int \sigma ^{\cdot }\overline{\sigma }^{\cdot }du$ or 
\begin{equation}
P_{a}^{\cdot }(u_{B})=-\int \sigma ^{\cdot }\overline{\sigma }^{\cdot }%
\widehat{l}_{a}dS.  \label{evolution}
\end{equation}

If $\psi _{2}^{*0},$ from Eq.(\ref{3}), is substituted into (\ref{twisting2*}
) we obtain after using Eq.(\ref{1}) and regrouping, a truly ugly equation
that we write as

\begin{equation}
\psi _{1}^{*0\,\cdot }-\text{ $\frak{d}$}^{3}\overline{\sigma }+\text{ $%
\frak{d}$}\Psi +3L^{\cdot }\Psi =(NL)  \label{psi0dot}
\end{equation}
with the non-linear terms, (NL), expressed by

\begin{eqnarray}
(NL) &\equiv &\text{ $\frak{d}$}[\overline{\sigma }^{\cdot }\sigma ]+2\sigma 
\text{  $\frak{d}$}\overline{\sigma }^{\cdot }+3L\text{ $\frak{d}$}^{2}%
\overline{\sigma }+L^{3}\overline{\sigma }^{\cdot \cdot \cdot }+3L\sigma 
\overline{\sigma }^{\cdot \cdot }  \label{NL} \\
&&+3L^{2}\text{ $\frak{d}$}\overline{\sigma }^{\cdot \cdot }+3\dot{L}[2L%
\text{ $\frak{d}$}\overline{\sigma }^{\cdot }+L^{2}\overline{\sigma }^{\cdot 
\cdot }+\text{ $\frak{d}$}^{2}\overline{\sigma }+\sigma \overline{\sigma }^{%
\cdot }].  \nonumber
\end{eqnarray}

The grouping has the non-linear higher order terms put into (NL) while the
`controllable' (lower order) terms are on the left side. It is from this
equation and (\ref{evolution}) we find the equations of motion. We first
require, as we said earlier, that $\psi _{1}^{*0\,}$ (and of course $\psi
_{1}^{*0\cdot \,}$) have no $l=1$ harmonics. From the fact that  $\frak{d}%
^{3}\overline{\sigma },$ from its spin $s=2$ structure, has no $l=1$ terms
we see that Eq.(\ref{psi0dot}) reduces to the condition that 
\begin{equation}
(\text{ $\frak{d}$}\Psi +3L^{\cdot }\Psi )_{l=1}=(NL)_{l=1}.
\label{condition}
\end{equation}
It is this equation that we must analyze, or more accurately approximate, by
looking only at the leading spherical harmonic terms. [See the following
subsection for an alternative derivation of Eq.(\ref{condition}).]

From Eqs.(\ref{PSI.II} ) and (\ref{L(u)}), considering only the lowest order 
$l=1$ terms we have 
\begin{eqnarray*}
\Psi (u_{B},\zeta ,\overline{\zeta }) &=&-2\frac{G}{c^{2}}(\sqrt{2}%
M-3P^{i}c_{i}) \\
L(u_{B},\zeta ,\overline{\zeta }) &=&\xi ^{a}(\tau )\widehat{m}_{a}(\zeta ,%
\overline{\zeta })
\end{eqnarray*}
from which it follows that

\begin{eqnarray}
\text{ $\frak{d}$}\Psi  &=&12\frac{G}{c^{2}}P^{a}\widehat{m}_{a}
\label{inter} \\
3L^{\cdot }\Psi  &=&-6\frac{G}{c^{2}}\xi ^{a\,\cdot }(\tau )\widehat{m}_{a}\{%
\sqrt{2}M-3P^{i}c_{i}\} \\
&=&-\sqrt{2}6\frac{G}{c^{2}}M\xi ^{a\,\cdot }(\tau )\widehat{m}_{a}+18\frac{G%
}{c^{2}}\xi ^{a\,\cdot }P^{b}(\tau )\widehat{m}_{a}\widehat{c}_{b} \\
&=&-\sqrt{2}6\frac{G}{c^{2}}M\xi ^{a\,\cdot }(\tau )\widehat{m}_{a}+9i\frac{G%
}{c^{2}}\xi ^{a\,\cdot }P^{b}(\tau )\epsilon _{abcd}\widehat{m}%
^{c}t^{d}+Y\{l\eqslantgtr 2\}
\end{eqnarray}
where we have used the Clebsch-Gordon expansion [see Appendix C] of the
product $\widehat{m}_{a}c_{b}$. From Eq.(\ref{condition}), we then have 
\[
12\frac{G}{c^{2}}P^{a}\widehat{m}_{a}-\sqrt{2}6\frac{G}{c^{2}}M\xi ^{a\,%
\cdot }(\tau )\widehat{m}_{a}+9i\frac{G}{c^{2}}\xi ^{a\,\cdot }P^{b}(\tau
)\epsilon _{abcd}\eta ^{cf}\widehat{m}_{f}t^{d}=(NL)_{l=1}\equiv GN^{a}%
\widehat{m}_{a}
\]
or, using the rescaled $u_{B}$ to the conventional Bondi $u_{c}$ coordinate%
\ref{convention}\cite{coorTransf}, by $u_{B}=\frac{u_{c}}{\sqrt{2}}$.

\begin{equation}
P^{e}=M\xi ^{e\,\cdot }(\tau )-i\frac{3\sqrt{2}}{4}\xi ^{a\,\cdot
}P^{b}(\tau )\epsilon _{abcd}\eta ^{ce}t^{d}+D^{e}.  \label{3-momentum}
\end{equation}
We have the three-momentum expressed in terms of the mass $M$ and the
complex velocity, $\xi ^{a\,\cdot }.$ Remembering, from Eq.(\ref{twisting8*}%
) that $\Psi $ is real we see, from Eq.(\ref{PSI.II}), that $M$ and $P^{c}$
are real. By defining the real and imaginary parts of $\xi ^{a\,\cdot },$%
\begin{equation}
\xi ^{a\,\cdot }=\xi _{R}^{a\,\cdot }+i\xi _{I}^{a\,\cdot },
\label{splitting}
\end{equation}
and taking the real and imaginary parts of (\ref{3-momentum}), (ignoring $%
D^{f}$ in this approximation)

\begin{eqnarray}
P^{e} &=&M\xi _{R}^{e\,\cdot }(\tau )+\frac{3\sqrt{2}}{4}\xi _{I}^{a\cdot
}P^{b}(\tau )\epsilon _{abcd}\eta ^{ce}t^{d}  \label{couplingI} \\
M\xi _{I}^{e\,\cdot }(\tau ) &=&\frac{3}{4}\xi _{R}^{a\,\cdot }P^{b}(\tau
)\epsilon _{abcd}\eta ^{ce}t^{d}.  \label{couplingII}
\end{eqnarray}

If we identify the spin-angular momentum (from the analogy with the Kerr
metric) by 
\[
S^{b}=M\xi _{I}^{b} 
\]
and $\xi _{R}^{b\,\cdot }$ $\equiv v^{b}$ with a real velocity vector and
linearize, we have

\begin{eqnarray*}
P^{c} &=&Mv^{c} \\
S^{b\,\cdot \,} &=&\frac{M^{\,\cdot }}{M}S^{b}
\end{eqnarray*}
which, with 
\[
P_{a}^{\cdot }(u_{B})=(Mv^{c})^{\cdot }=-\int \sigma ^{\cdot }\overline{%
\sigma }^{\cdot }\widehat{l}_{a}dS, 
\]
yields the equations of motion for the center of mass and spin-angular
momentum - the equations of motion of a spinning particle with mass and
momentum loss via radiation $-$ all derived from the Bianchi identities and
driven by the original Bondi shear.

\begin{remark}
Note that from the linearization of Eqs.(\ref{couplingI}) and (\ref
{couplingII}), the coupling between the spin and the velocity vector has
disappeared. The details of the coupling will be investigated in a future
note.
\end{remark}

There is an important point to be made here. We know that the complex curve
we wish to determine is given by the expression 
\[
z^{a}=\xi ^{b}(\tau ), 
\]
with the complex paramter $\tau ,$ while in the above expressions for the
determination of the curve we have treated $\xi ^{b}$ to be a complex
funtion of the real parameter $u$. This is justified by realizing that we
have been making severe approximations, throwing out higher non-linear
terms, not using more and higher Clebsch-Gordon expansions and from the
linearization of the inversion of $\tau =T(u,\zeta ,\overline{\zeta }),$
Appendix A. In the lowest order of the inversion we do have $\tau =u.$

The basic idea we are espousing is that there is, in $H$-Space, a unique
complex curve that can be called the intrinsic complex center of mass
world-line. It is determined from \textit{a geometric structure on the
physical space-time, the null direction field,} independent of the choice of
coordinates on Scri. In order to obtain - as an approximation - what one
would normally call the orbital and spin angular momentum in linear theory
one must use Eq.(\ref{BtoT2}) and extract by an integration, from the Bondi $%
\psi _{1}^{0},$ its $l=1$ coefficient. The values of this `normally' defined
angular momentum will, in general, depended on the choice of the scri
coordinates or the Bondi frame.

\begin{remark}
If we linearize (decouple) Eq.(\ref{3-momentum}) we have 
\[
P^{e}=M\xi ^{e\,\cdot }(u)
\]
so that if we treat $M$ and $P^{e}$ as very slowly varying we can integrate
it as 
\[
\xi ^{e\,}=\xi _{R}^{a\,}+i\xi _{I}^{a\,}=\frac{P^{e}}{M}u
\]
or, since $M$ and $P^{e}$ are real, 
\begin{eqnarray*}
\xi _{R}^{a\,}=\frac{P^{e}}{M}u \\
\xi _{I}^{a\,}=const=\text{initial conditions,}
\end{eqnarray*}
thus obtaining the usual motion for the center of mass.
\end{remark}

\subsection{\textbf{An Alternative Route to the Complex Curve}}

There is an alternative method of finding the curve that in some sense is
more direct than the one just given.

We begin by inverting Eqs.(\ref{BondiToTwist}) - (\ref{BtoT5}) and then
writing 
\[
\psi _{1}^{*0}=\psi _{1}^{0}-3L\psi _{2}^{0}+3L^{2}\psi _{3}^{0}-L^{3}\psi
_{4}^{0} 
\]
and its $u_{B}$-derivative 
\begin{equation}
\psi _{1}^{*0\,\cdot }=(\psi _{1}^{0}-3L\psi _{2}^{0}+3L^{2}\psi
_{3}^{0}-L^{3}\psi _{4}^{0})^{\,\cdot }  \label{psidot}
\end{equation}

We then impose our basic requirement for the determination of the curve,
namely that the lowest harmonic, i.e., the $l=1$ term in $\psi _{1}^{*0\,}$%
and $\psi _{1}^{*0\,\cdot }$ should vanish. Eq.(\ref{psidot}) can then be
written as

\begin{equation}
\{\psi _{1}^{0\,\cdot }\}_{l=1}=\{(3L\psi _{2}^{0}-3L^{2}\psi
_{3}^{0}+L^{3}\psi _{4}^{0})^{\,\cdot }\}_{l=1}+[\text{Terms]}_{l\geq 2}.
\label{psidot2}
\end{equation}
Eq.(\ref{psidot2}), using the Bianchi identities for $\psi _{1}^{0\,\cdot }$
and $\psi _{1}^{0\,\cdot }$ and expressions $\psi _{3}^{0}\ $and $\psi
_{4}^{0},$ i.e., Eqs.(\ref{Bondi2}) - (\ref{Bondi5}), again leads to the
complex curve by the following argument.

Using 
\begin{equation}
\Psi =\psi _{2}^{0\,}+\text{ $\frak{d}$}^{2}\overline{\sigma }+\sigma 
\overline{\sigma }^{\cdot }  \label{Mass Aspect2}
\end{equation}
instead of $\psi _{2}^{0\,}$ and the evolution equation for $\Psi ,$ Eq.(\ref
{BondiMassLoss}), (really a Bianchi identity)

\begin{equation}
\Psi ^{\cdot }=\sigma ^{\cdot }\overline{\sigma }^{\cdot }
\label{BondiMassLoss2}
\end{equation}
Eq.(\ref{psidot2}) can be rewritten as

\begin{equation}
\{3L^{\cdot }\text{ }\Psi -\psi _{1}^{0\,\cdot }\}_{l=1}=NL_{l=1}+[\text{
Terms]}_{l\geqslant 2},  \label{psidot3}
\end{equation}
while the Bianchi identity for $\psi _{1}^{0\,}{}^{\cdot }$ becomes 
\begin{equation}
\psi _{1}^{0\,\cdot }=-\text{ $\frak{d}$}\Psi +NL+[\text{Terms]}_{l\geqslant
2}.  \label{bianchi2}
\end{equation}

In both cases, for our approximation, we have simply grouped the higher
harmonic terms and the nasty non-linear terms into $[$Terms]$_{l\geqslant 2}$
and $NL$.

By substituting the $\psi _{1}^{0\,\cdot }$ from Eq.(\ref{bianchi2}) into (%
\ref{psidot3}) we obtain 
\begin{equation}
(\text{ $\frak{d}$}\Psi +3L^{\cdot }\Psi )_{l=1}=(NL)_{l=1},  \label{curve2}
\end{equation}
which is identical to Eq.(\ref{condition}). This relationship, with Eqs.(\ref
{L(u)}) and (\ref{u2}),

\begin{eqnarray}
L(u_{B},\zeta ,\overline{\zeta }) &=&\xi ^{a}(\tau )\widehat{m}_{a}(\zeta ,%
\overline{\zeta })+\text{ $\frak{d}$}_{(\tau )}X_{(l\geqslant 2)}(\tau
,\zeta ,\overline{\zeta })  \label{L(u)2} \\
u_{B} &=&\xi ^{a}(\tau )\widehat{l}_{a}(\zeta ,\overline{\zeta }%
)+X_{(l\geqslant 2)}(\tau ,\zeta ,\overline{\zeta })  \label{u22}
\end{eqnarray}
leads, via the argument in the previous section, to the equations for the
complex curve.

\section{Examples}

\subsection{The Robinson-Trautman metrics}

We begin with the Bianchi Identities Eqs.(\ref{NU1}) - (\ref{NU6}), which
are essentially the same as Eqs.(\ref{twisting1})-(\ref{twisting8*}),
differing only by the coordinate transformation $\tau \Rightarrow u_{B},$
given by $u_{B}=X(\tau ,\zeta ,\overline{\zeta }),$ from one set to the
other. The tetrad system, as well as the stereographic angle $L,$ are the
same with both sets. In other words, geometrically, they represent the same
situation. The $L$ or the $l^{*a}$ describe a surface forming congruence. We
imposed, in the last section, the condition that the $l=1$ harmonics of the
tetrad component, $\psi _{1}^{0},$ vanish. In this example we simply
strengthen this $condition$ by requiring that 
\[
\psi _{1}^{0}=0.
\]
This condition, from Eq.(\ref{NU2}) forces $\psi _{2}^{0}$ to be a function
only of $\tau ,$ i.e., $\psi _{2}^{0}=\psi _{2}^{0}(\tau )=2\sqrt{2}GM(\tau
)/c^{2},$ so that Eq.(\ref{NU3}) becomes 
\begin{equation}
\psi _{2}^{**0\,\prime }=-\overline{\frak{d}}\text{$\frak{d}$}\overline{%
\frak{d}}\text{$\frak{d}$}\log V+3\frac{V^{\prime }}{V}\psi _{2}^{**0}.
\label{RT1}
\end{equation}
(The strange numerical factor $2\sqrt{2}$ arises from our use of the radial
coordinate $r$ that differs from the conventional one by the factor$\sqrt{2}.
$)

By the reparametrization freedom, Eq.(\ref{repar}), a change in the $\tau $
coordinate can be made, i.e., $\tau =F(\tau ^{*}),$ so that the $\psi
_{2}^{**0\,}$ is constant, thus yielding

\begin{equation}
\frac{V^{\prime }}{V}\psi _{2}^{**0}=\overline{\frak{d}}\text{$\frak{d}$}%
\overline{\frak{d}}\text{$\frak{d}$}\log V  \label{RT2}
\end{equation}
the well-known Robinson-Trautman equation\cite{RT} describing \textit{an
asymptotic} Robinson-Trautman space-time. If we further assumed that $\psi
_{0}=0$ this would lead to the Robinson-Trautman metric.

Note that since 
\begin{eqnarray*}
u &=&X=\xi ^{a}(\tau )\widehat{l}_{a}(\zeta ,\overline{\zeta }
)+X_{(l\geqslant 2)}(\tau ,\zeta ,\overline{\zeta }) \\
V &=&X^{\prime }=\xi ^{a\,\prime }(\tau )\widehat{l}_{a}(\zeta ,\overline{%
\zeta })+X_{(l\geqslant 2)}^{\prime \,}(\tau ,\zeta ,\overline{\zeta })
\end{eqnarray*}
we see that Eq.(\ref{RT2}), is a differential equation for $V$ and more
specifically the coefficients of the $l=0$ $\&$ $1$ harmonics have the form
of Newton's second law of motion 
\[
M\xi ^{a\,\prime \prime }=F^{a} 
\]
so that the Robinson-Trautman space-time describes a gravitational rocket
ship.

If we do have a solution to the Robinson-Trautman equation, i.e., $V=V(\tau
,\zeta ,\overline{\zeta }),$ the associated Bondi shear can be easily
constructed from the implicit relations 
\begin{eqnarray}
u_{B} &=&X(\tau ,\zeta ,\overline{\zeta })=\int V(\tau ,\zeta ,\overline{%
\zeta })d\tau   \label{BS} \\
\sigma (u_{B},\zeta ,\overline{\zeta }) &=&\text{ $\frak{d}$}_{(\tau
)}^{2}X(\tau ,\zeta ,\overline{\zeta }).  \nonumber
\end{eqnarray}

\subsection{Twisting Type II Metrics}

We can describe a more general situation by going to Eqs.(\ref{twisting1})-(%
\ref{twisting8*}) and again require that the Weyl components 
\begin{eqnarray*}
\psi _{1}^{*0} &=&0 \\
\psi _{0}^{*0\,} &=&0.
\end{eqnarray*}
This leads to the equations

\begin{eqnarray}
\text{ $\frak{d}$}\psi _{2}^{*0}+L\psi _{2}^{*0\,\cdot }+3L^{\cdot }\psi
_{2}^{*0} &=&0  \label{typeIIa} \\
\text{ $\frak{d}$}\psi _{3}^{*0}+L\,\psi _{3}^{*0\,\cdot }+2L^{\cdot }\psi
_{3}^{*0} &=&-\psi _{2}^{*0\,\cdot },  \label{typeIIb}
\end{eqnarray}
with

\begin{eqnarray}
\psi _{3}^{*0} &=&\text{ $\frak{d}$}\overline{\sigma }^{\cdot }+L\overline{%
\sigma }^{\cdot \cdot },  \label{typeIIc} \\
\sigma  &\equiv &\text{ $\frak{d}$}L+\frac{1}{2}(L^{2})^{\cdot },  \nonumber
\\
\Psi  &=&\psi _{2}^{*0}+2L\text{ $\frak{d}$}\overline{\sigma }^{\cdot }+L^{2}%
\overline{\sigma }^{\cdot \cdot }+\text{ $\frak{d}$}^{2}\overline{\sigma }%
+\sigma \overline{\sigma }^{\cdot },  \nonumber \\
\Psi -\overline{\Psi } &=&0,  \nonumber
\end{eqnarray}
that are precisely the equations for the twisting algebraically special type
II metrics. A detailed analysis of these equations and their generalization
to the type II Einstein-Maxwell equations is in preparation.

\subsection{The Sparling-Tod Shear}

George Sparling and Paul Tod\cite{TS} found a class of Bondi shears such
that the good-cut equation (\ref{GoodCut}) 
\begin{equation}
\text{ $\frak{d}$}^{{2}}X=\sigma (X,\zeta ,\overline{\zeta })  \label{GC}
\end{equation}
can be integrated exactly. An example of their class is 
\begin{equation}
\sigma (u,\zeta ,\overline{\zeta })=\frac{(q_{a}m^{a})^{2}}{u^{3}}
\label{ST}
\end{equation}
with $q_{a}$ the constant vector 
\begin{equation}
q_{a}=(0,q_{i}).  \label{q}
\end{equation}

The solution to (\ref{GC}) can be given by

\begin{equation}
u^{2}=X^{2}(z^{a},\zeta ,\overline{\zeta })=Z^{2}+S^{2}  \label{STsol}
\end{equation}
with 
\begin{eqnarray}
Z &=&z^{a}l_{a}(\zeta ,\overline{\zeta }),  \label{ZS} \\
S &=&s^{a}l_{a}(\zeta ,\overline{\zeta }).  \nonumber
\end{eqnarray}

The four $z^{a}=(t,x,y,z)$ are the $H$-Space coordinates (four constants of
integration) and the $s^{a}(z^{b})$ are four functions of $z^{a}$ determined
by the algebraic equation

\begin{equation}
s^{a}z^{b}(l_{a}m_{b}-l_{b}m_{a})=q_{a}m^{a}.  \label{algeq}
\end{equation}

Using the identity (see Appendix C for the notation and conventions)

\begin{equation}
(\widehat{l}_{a}\widehat{m}_{b}-\widehat{m}_{a}\widehat{l}_{b})=\frac{1}{2}(%
\widehat{t}_{a}\widehat{m}_{b}-\widehat{m}_{a}\widehat{t}_{b})-\frac{1}{2}%
i\epsilon _{abcd}\widehat{m}^{c}t^{d}  \label{Identity}
\end{equation}
and the ansatz

\begin{eqnarray}
s^{a} &=&S_{0}\text{ }t^{a}+S_{1}\text{ }q^{a}+i\text{ }S_{2}\sqrt{2}\text{ }
\varepsilon ^{a}{}_{bc}q^{b}z^{c}  \label{ansatz} \\
\sqrt{2}\text{ }\varepsilon ^{a}{}_{bc} &=&-\text{ }\varepsilon
^{a}{}_{bcd}t^{d}
\end{eqnarray}
one find that the ($S_{0},S_{1},S_{2}$) are uniquely determined so that

\begin{equation}
s^{a}=\frac{1}{z_{e}z^{e}}\{z_{b}q^{b}t^{a}-z^{b}t_{b}q^{a}+i\text{ }%
\varepsilon ^{a}{}_{bcd}t^{d}q^{b}z^{c}\}.  \label{sa}
\end{equation}

Note that though $s^{a},$\ in Eq.(\ref{algeq}) was defined up to a term
proportional to $z^{a}$, the ansatz determined it uniquely.

Eqs.(\ref{sa}), (\ref{STsol}) and (\ref{SZ}) determine the solution $%
u_{B}=X(z^{b},\zeta ,\overline{\zeta })$ so when an $H$-Space world-line, $%
z^{b}=\xi ^{b}(\tau ),$ is chosen we have the null direction field

\begin{eqnarray}
L(u_{B},\zeta ,\overline{\zeta }) &=&\text{ $\frak{d}$}_{(\tau )}X(\xi
^{a}(\tau ),\zeta ,\overline{\zeta })=\frac{Z\text{ $\frak{d}$}Z+S\text{ $%
\frak{d}$}S}{\sqrt{Z^{2}+S^{2}}}  \label{STL} \\
u_{B} &=&\sqrt{Z^{2}+S^{2}}.  \nonumber
\end{eqnarray}

\subsection{Arbitrary Construction}

For further general examples, we can reverse the process of first giving the
shear and then finding the $L(u_{B},\zeta ,\overline{\zeta })$. Instead, we
can choose an arbitrary, $s=0,$ function with the form 
\[
u_{B}=X(\tau ,\zeta ,\overline{\zeta })=\xi ^{a}(\tau )\widehat{l}_{a}(\zeta
,\overline{\zeta })+X_{(l\geqslant 2)}(\tau ,\zeta ,\overline{\zeta })
\]
and then \textit{define} 
\[
L(u_{B},\zeta ,\overline{\zeta })=\text{ $\frak{d}$}X=\xi ^{a}(\tau )%
\widehat{m}_{a}(\zeta ,\overline{\zeta })+\text{ $\frak{d}$}X_{(l\geqslant
2)}(\tau ,\zeta ,\overline{\zeta }).
\]
The associated Bondi shear is then given, implicitly, by 
\begin{eqnarray*}
\sigma (u_{B},\zeta ,\overline{\zeta }) &=&\text{ $\frak{d}$}%
^{2}X_{(l\geqslant 2)}(\tau ,\zeta ,\overline{\zeta }) \\
u &=&\xi ^{a}(\tau )\widehat{l}_{a}(\zeta ,\overline{\zeta })+X_{(l\geqslant
2)}(\tau ,\zeta ,\overline{\zeta }).
\end{eqnarray*}

\section{Conclusion}

In this work we have shown that for all asymptotically flat space-times
there is a hidden structure that must be extracted dynamically from the
known asymptotic Weyl tensor and its related characteristic data, the Bondi
shear, $\sigma $. This hidden structure is a specific field of
asymptotically shear-free null directions - or an asymptotically shear-free
null geodesic congruence. Within the information for the description of this
direction field is a complex world-line that is defined in the $H$-space
associated with the given asymptotically flat space-time. One can try to
give meaning to this world-line by \textit{defining} it to be the \textit{\
complex center of mass }of the interior gravitating system. Though there is
no proof or even overwhelming evidence that this assignment is completely
reasonable, we can ask the question: what physical justification can be
given for this assignment. On the surface it certainly is strange - where
does a complex world-line in $H$-space enter in any direct physical
observation.

We would like to try to give a justification by three different types of
argument.

1. Examples: In several known cases, e.g., the Kerr-metric, the charged-Kerr
metric, the Lienard-Wiechert-Maxwell fields, the Kerr-Maxwell field, all
have a center of mass (or center of charge in the Maxwell case) that is
identical to the more general construction described here. In addition, the
construction given here coincides with the definitions from linearized
theory.

2. Though it appears to us to be surprising and remarkable, one does get, in
a first approximation, the conventional equations of motion of a complicated
spinning gravitating source directly from the asymptotic information - with
no details needed about the interior source. These equations contain the
radiation reaction from the gravitational radiation that arises from the
Bondi mass loss.

3. It is possible to ask the question: can one give, at least in principle,
an observational means of ``observing'' this complex motion? The answer, we
believe, is yes - though to really do so is impossible. It involves having a
huge number of observers surrounding the gravitating source and looking at
all the null rays reaching them, then picking out shear-free null direction
fields and by angular integration finding the complex world-line. This type
of argument must be tightened and made more precise.

Work has begun on the application of these ideas to asymptotically flat
Einstein-Maxwell fields. We remark that it appears (in a preliminary stage)
that there are two complex world-lines, a complex center of mass obtained
from the asymptotic Weyl tensor and a complex center of charge obtained form
the asymptotic Maxwell field. Their imaginary parts give respectively the
spin and magnetic dipole moments. If the two world-lines coincide then the
Dirac value of gyromagnetic ratio follows\cite{gyro}.

Many years ago, by a rather strange and not at all understood complex
coordinate transformation applied to the Reissner-Nordstrom Einstein-Maxwell
field, the charged Kerr Einstein-Maxwell field was first found\cite{QKerr}.
The work described here appears to clarify and even explain this mysterious
transformation.

Nevertheless, there is still much to be understood about the present
material, i.e., the asymptotic shear-free congruences and the complex
H--space curves. Is it just an coincidence that it appears to be so
connected with physical issues or does something deeper lie behind it?

\section{Acknowledgments}

This material is based upon work (partially) supported by the National
Science Foundation under Grant No. PHY-0244513. Any opinions, findings, and
conclusions or recommendations expressed in this material are those of the
authors and do not necessarily reflect the views of the National Science.
E.T.N. thanks the NSF for this support. C.N.K. would like to thank CONICET
for support. G.S.O. acknowledges the financial support from CONACyT through
Grand No.44515-F, VIEP-BUAP through Grant No. II161-04/EXA and Sistema
Nacional de Investigadores (SNI-M\TEXTsymbol{\backslash}'exico).

\section{Appendix A}

In order to obtain $\tau =T(u_{B},\zeta ,\overline{\zeta }),$ the inversion
of 
\[
u_{B}=X(\tau ,\zeta ,\overline{\zeta })=z^{a}(\tau )\widehat{l}_{a}(\zeta ,%
\overline{\zeta })+X_{(l\geqslant 2)}(\tau ,\zeta ,\overline{\zeta }), 
\]
we first note that $\tau $ can be replaced by any function of $\tau ,$ i.e.,
the Eq.(\ref{L1}) is invariant under $\widehat{\tau }=F(\tau ).$ See Eq.(\ref
{repar}). Thus by taking 
\[
z^{0}(\tau )=\tau 
\]
we have 
\[
u_{B}=\tau +z^{i}(\tau )\widehat{l}_{i}(\zeta ,\overline{\zeta }%
)+X_{(l\geqslant 2)}(\tau ,\zeta ,\overline{\zeta }) 
\]
or 
\[
\tau =u_{B}-z^{i}(\tau )\widehat{l}_{i}(\zeta ,\overline{\zeta }%
)+X_{(l\geqslant 2)}(\tau ,\zeta ,\overline{\zeta }). 
\]
This can be solved by iteration to any order: 
\begin{eqnarray*}
\tau _{0} &=&u_{B} \\
\tau _{1} &=&u_{B}-z^{i}(u_{B})\widehat{l}_{i}(\zeta ,\overline{\zeta }%
)+X_{(l\geqslant 2)}(u_{B},\zeta ,\overline{\zeta }) \\
\tau _{n} &=&u_{B}-z^{i}(\tau _{n-1})\widehat{l}_{i}(\zeta ,\overline{\zeta }%
)+X_{(l\geqslant 2)}(\tau _{n-1},\zeta ,\overline{\zeta }).
\end{eqnarray*}

\section{Appendix B}

Normalization and conventions: Our goal here is to express the function $%
\Psi $ in terms of the Bondi mass/momenta and also show how to extract them
from the $\Psi .$

We start, in a given frame in Minkowski space, with $C_{a}$ a unit radial
space-like vector and $T_{a}$ a unit time-like vector given by

\begin{eqnarray}
C^{a} &=&(0,\cos \phi \sin \theta ,\sin \phi \sin \theta ,\cos \theta
)=-C_{a}, \\
T_{a} &=&(1,0,0,0).
\end{eqnarray}
with the null vectors 
\begin{eqnarray}
\widehat{l}_{a} &=&\frac{1}{\sqrt{2}}\left( 1,-\cos \phi \sin \theta ,-\sin
\phi \sin \theta ,-\cos \theta \right) =\frac{1}{\sqrt{2}}(T_{a}+C_{a}) \\
\widehat{n}_{a} &=&\frac{1}{\sqrt{2}}\left( 1,\cos \phi \sin \theta ,\sin
\phi \sin \theta ,\cos \theta \right) =\frac{1}{\sqrt{2}}(T_{a}-C_{a})
\end{eqnarray}
and 
\[
\widehat{c}_{a}=\widehat{l}_{a}-\widehat{n}_{a}=-\sqrt{2}(0,\cos \phi \sin
\theta ,\sin \phi \sin \theta ,\cos \theta )=\sqrt{2}C_{a} 
\]

We let $a=(0,1,2,3)=(0,i)$ and $\widehat{l}_{a}=(l_{0},l_{i})$ and $\widehat{%
c}_{i}=\sqrt{2}C_{i}$, etc.

We then define the $l=0$ and $l=1$ parts of the mass aspect $\Psi $ by

\begin{eqnarray}
\Psi  &\equiv &\psi _{2}^{*0}+2L\text{ $\frak{d}$}\overline{\sigma }^{\cdot %
}+L^{2}\overline{\sigma }^{\cdot \cdot }+\text{ $\frak{d}$}^{2}\overline{%
\sigma }+\sigma \overline{\sigma }^{\cdot }  \label{PSI1} \\
\Psi  &=&\chi -\chi ^{i}\widehat{c}_{i}+....  \label{PSI2}
\end{eqnarray}
Therefore 
\begin{eqnarray}
\Psi \widehat{l}_{a} &=&\chi \widehat{l}_{a}-\chi ^{i}\widehat{c}_{i}%
\widehat{l}_{a}+...  \label{PSIL} \\
&=&\frac{1}{\sqrt{2}}[\chi T_{a}-\sqrt{2}\chi ^{i}C_{i}C_{a}+\chi C_{a}-%
\sqrt{2}\chi ^{i}C_{i}T_{a}].  \nonumber
\end{eqnarray}
From the integral identities 
\begin{eqnarray*}
\int dS &=&4\pi  \\
\int C_{i}dS &=&0 \\
\int C_{i}C_{j}dS &=&\frac{4\pi }{3}\delta _{ij}
\end{eqnarray*}
we have that

\begin{eqnarray}
\int \Psi \widehat{l}_{a}dS &=&\frac{1}{\sqrt{2}}\int dS(\chi T_{a}-\sqrt{2}
\chi ^{i}C_{i}C_{a})  \nonumber \\
&=&\frac{4\pi }{\sqrt{2}}[\chi T_{a}-\chi ^{i}\frac{\sqrt{2}}{3}]=\frac{4\pi 
}{\sqrt{2}}\left( \chi ,\chi _{i}\frac{\sqrt{2}}{3}\right) .  \label{intPsi}
\end{eqnarray}
with $\chi _{i}=-\chi ^{i}.$\newline

On the other hand, from reference (\cite{TN})\textit{\ }and a change of%
\textit{\ }notation\cite{*coordinate change},\textit{\ \ }we have that the
4-momentum for an asymptotically flat space-time is defined by

\begin{eqnarray}
P_{a} &=&-\frac{c^{2}}{8\pi G}\int \Psi \widehat{l}_{a}dS,  \label{P_a} \\
\Psi &=&-2\sqrt{2}\frac{G}{c^{2}}(M+.....
\end{eqnarray}
then from Eqs.(\ref{P_a}),(\ref{intPsi}) and (\ref{PSIL}),

\begin{eqnarray}
P_{a} &=&-\frac{c^{2}}{8\pi G}[\frac{4\pi }{\sqrt{2}}\left( \chi ,\chi _{i}%
\frac{\sqrt{2}}{3}\right) ]  \label{P_a*} \\
(P_{0},P_{i}) &=&(M,P_{i})=-\frac{c^{2}}{2\sqrt{2}G}\left( \chi ,\chi _{i}%
\frac{\sqrt{2}}{3}\right) .  \nonumber
\end{eqnarray}

We find that: 
\begin{eqnarray}
\chi &=&-2\sqrt{2}\frac{G}{c^{2}}M  \label{chi} \\
\chi _{i} &=&-6GP_{i}=6\frac{G}{c^{2}}P^{i}  \label{chii}
\end{eqnarray}

and therefore we have 
\begin{equation}
\Psi =\chi -\chi ^{i}\widehat{c}_{i}+....=-2\frac{G}{c^{2}}(\sqrt{2}M-3P^{i}%
\widehat{c}_{i})+....
\end{equation}

\section{Appendix C}

Relations between the following quantities have been used through this work:

\begin{eqnarray}
\widehat{l}^{a} &=&\frac{\sqrt{2}}{2}(1,\frac{\zeta +\overline{\zeta }}{%
1+\zeta \overline{\zeta }},-i\frac{\zeta -\overline{\zeta }}{1+\zeta 
\overline{\zeta }},\frac{-1+\zeta \overline{\zeta }}{1+\zeta \overline{\zeta 
}}),  \label{tetrad} \\
\widehat{m}^{a} &=&\text{ $\frak{d}$}\widehat{l}^{a}=\frac{\sqrt{2}}{2}(0,%
\frac{1-\overline{\zeta }^{2}}{1+\zeta \overline{\zeta }},\frac{-i(1+%
\overline{\zeta }^{2})}{1+\zeta \overline{\zeta }},\frac{2\overline{\zeta }}{%
1+\zeta \overline{\zeta }}),  \nonumber \\
\widehat{\overline{m}}^{a} &=&\overline{\frak{d}}\widehat{l}^{a}=\frac{\sqrt{%
2}}{2}(0,\frac{1-\zeta ^{2}}{1+\zeta \overline{\zeta }},\frac{i(1+\zeta ^{2})%
}{1+\zeta \overline{\zeta }},\frac{2\zeta }{1+\zeta \overline{\zeta }}), 
\nonumber \\
\widehat{t}^{a} &=&\sqrt{2}(1,0,0,0),  \nonumber \\
\widehat{n}^{a} &=&\widehat{t}^{a}-\widehat{l}^{a}=\frac{\sqrt{2}}{2}(1,-%
\frac{\zeta +\overline{\zeta }}{1+\zeta \overline{\zeta }},i\frac{\zeta -%
\overline{\zeta }}{1+\zeta \overline{\zeta }},\frac{1-\zeta \overline{\zeta }%
}{1+\zeta \overline{\zeta }}),  \nonumber \\
\widehat{c}^{a} &=&\widehat{l}^{a}-\widehat{n}^{a}=\sqrt{2}(0,\frac{\zeta +%
\overline{\zeta }}{1+\zeta \overline{\zeta }},-i\frac{\zeta -\overline{\zeta 
}}{1+\zeta \overline{\zeta }},\frac{-1+\zeta \overline{\zeta }}{1+\zeta 
\overline{\zeta }}),  \nonumber
\end{eqnarray}

with the products

\begin{eqnarray*}
(\widehat{m}_{a}\widehat{\overline{m}}_{b}-\widehat{\overline{m}}_{a}%
\widehat{m}_{b}) &=&-\frac{i}{2}\epsilon _{abcd}\widehat{c}^{c}t^{d}, \\
(\widehat{m}_{a}\widehat{c}_{b}-\widehat{c}_{a}\widehat{m}_{b}) &=&i\epsilon
_{abcd}\widehat{m}^{c}t^{d}. \\
\widehat{m}_{a}\widehat{c}_{b} &=&\frac{1}{2}(\widehat{m}_{a}\widehat{c}_{b}-%
\widehat{m}_{b}\widehat{c}_{a})+\frac{1}{2}(\widehat{m}_{a}\widehat{c}_{b}+%
\widehat{m}_{b}\widehat{c}_{a}) \\
&=&i\frac{1}{2}\epsilon _{abcd}\widehat{m}^{c}t^{d}+(l=2)Terms
\end{eqnarray*}
and 
\[
\epsilon _{abc}=-\frac{1}{\sqrt{2}}\epsilon _{abcd}t^{d}. 
\]


\begin{thebibliography}{99}
\bibitem{coorTransf}  When the radial coordinate $r$ is rescaled by $\sqrt{2}%
,$ in order to keep the coordinate conditions unchanged, the $u$ coordinate
must be rescaled by $1/\sqrt{\text{2}}.$
\end{thebibliography}
\end{document}